# Weak dispersive forces between glass-gold macroscopic surfaces in alcohols

P.J. van Zwol, G. Palasantzas*, J. Th. M. DeHosson

Department of Applied Physics, Materials innovation institute M2i and Zernike Institute for

Advanced Materials

University of Groningen, 9747 AG Groningen, The Netherlands

**Abstract**

In this work we concentrate on an experimental validation of the Lifshitz theory for van der Waals and Casimir forces in gold-alcohol-glass systems. From this theory weak dispersive forces are predicted when the dielectric properties of the intervening medium become comparable to one of the interacting surfaces. Using inverse colloid probe atomic force microscopy dispersive forces were measured occasionally and under controlled conditions by addition of salt to screen the electrostatic double layer force if present. The dispersive force was found to be attractive, and an order of magnitude weaker than that in air. Although the theoretical description of the forces becomes less precise for these systems even with full knowledge of the dielectric properties, we find still our results in reasonable agreement with Lifshitz theory.

Pacs mumbers: 78.68.+m, 03.70.+k, 85.85.+j, 68.08.-p

___________________
Corresponding authors: e-mail: G.Palasantzas@rug.nl, petervanzwol@gmail.com

# I. INTRODUCTION

In 1948 H.B.G. Casimir predicted that the Quantum Electro Dynamics of the vacuum results in attraction of two parallel conducting surfaces [1]. This result was generalized by E. Lifshitz and coworkers as a general theory of van der Waals and Casimir dispersive forces between surfaces and intervening media with arbitrary dielectric functions [2]. Casimir's predictions in air and vacuum have been experimentally confirmed by a wide variety of measurements [3-5]. Very recently the theory has been experimentally confirmed also for ethanol as intervening medium only between metal surfaces [6]. Furthermore, the Lifshitz theory predicts vanishing dispersive forces when the dielectric function of the intervening medium is equal to the dielectric function of one surface over a wide frequency range of the electro-magnetic spectrum. Although in reality the dielectric functions will never exactly overlap over the whole frequency spectrum, when the dielectric properties are similar very small forces can be predicted. Sometimes these forces are accompanied with more exotic behavior such as a switch in sign at a particular distance as it was discussed in [7].

Precision testing of the Lifshitz theory in liquid solutions is not straightforward due to the additional presence of short ranged (a few nm) solvation, long ranged hydrodynamic, and electrostatic double layer forces [8]. The latter can be screened by adding salt to the solution, and the charging mechanism of surfaces in liquid electrolytes is discussed in literature [8]. Non-linear Poisson Boltzmann theory (with various boundary conditions and extensions such as charge regulation) describes most of the electrostatic phenomena in liquid solutions. The charge of a surface in liquid is thought to be related to the chemistry of the materials involved [8], which is usually changed by varying the acidity of the electrolyte.

Therefore, both dispersion and electrostatic forces will be considered in the present paper with focus on alcohols as the intervening medium, since their dielectric function is comparable to that of glass. Force measurements were performed using borosilicate glass (BSG) spheres since they are extremely smooth (typical rms roughness < 1 nm) allowing measurements down to a few nm separations.

## II. EXPERIMENTAL METHODS

### (a) Inverse colloid probe force microscopy

For the force measurements the inverse force sensing technique is used (Fig. 1), which was inspired by ref. [9] employed for future laboratory cosmology Casimir force measurements. In fact, the inverse force sensing technique has several advantages: *i)* the metal coatings on the cantilever, as applied by the manufacturer, do not appear to induce stress-bending of the lever, *ii)* sticking and coating spheres on a cantilever is time consuming and it is avoided in this way, *iii)* the substrate with spheres can survive sonic cleansing treatment with aggressive liquids (i.e., acetone), while a cantilever with a glued probe does not, *iv)* a lot of statistics can be obtained in a short time by measuring on multiple spheres. A disadvantage is that precise knowledge of the interaction area on the sphere is lacking.

All measurements were performed with the Picoforce AFM. A grounded gold coated cantilever (μmasch NSC12 tipless Cr-Au, stifness ~0.2-1N/m) acts as a plate. Many BSG spheres (Duke Scientific, amount of Silica >90% by weight) were placed on a soda lime glass substrate (Fig. 1). Thermal tuning is used to compute the spring constant $K$ of the cantilever [10, 11] because the hydrodynamic calibration of $K$ is less precise for our system since the error in the sphere radius R (~11%) and the error in the viscosity ($\eta$) measurement (~2-7% [6]) would lead to a ~23% error in $K$ (since $F_{hydro} \sim \eta R^2$) [11]. The uncertainty in the thermal tuning method (~10%) together with the uncertainty in sphere radius R of $\delta R \sim 11\%$ leads to a total force uncertainty of 15% ($\sim \sqrt{\delta R^2 + \delta K^2}$).

For testing purposes, thermal tuning is performed 16 times in air and liquid while varying the place of the laser spot over the cantilever, moving the cantilever from place to place and recalibrating the deflection sensitivity every time. In air we found K=0.42±0.02 N/m (Quality factor Q=101±1.5) and in liquid 0.40±0.04 N/m (Q=3.46±0.02) for the same cantilever. The length of the cantilever (L=300 μm) is much larger than the sphere diameter. During measurement the sphere is not exactly located at the end of the free cantilever, but is situated approximately L-0.5$R_{sph}$, which corresponds to a deviation of (1-290/300)x100 ~ 3.3%. We verified that the off-end loading effect was small by determining the deflection sensitivity with the cantilever end on a sphere, and on a flat plate. The difference was 7% which is within the error in K of the tuning method. We have measured the deflection sensitivity on multiple spheres and did not find large variations for the calibrated cantilever stiffness in accordance with [10]. The variations are in the range of the estimated accuracy of the tuning method. For example, if we change the lever in situ with another one, perform thermal tuning, and measure forces on the same sphere, the measured force does not change beyond the estimated error.

Notably, for the cantilevers used here (with 20 nm Cr and 20nm Au overall coating applied by the manufacturer) the linear signal was very small (<0.5nm over an 8 μm range) or not measurable. In general, linear signals are commonly observed effects when a cantilever is subject to stress-bending due to an applied coating.

The interaction of the cantilever with multiple spheres is avoided due to the short interaction range (<300 nm), and the fact that the cantilever makes an angle of *α~22º* with the surface. Furthermore, isolated spheres were used for the force measurements. While the tilt angle 'a' is included [10], tuning results in the inverse colloid probe case or the normal colloid probe (with a sphere on a cantilever) case are in principle the same (Fig. 2). This is because the cantilever deflection is calibrated from the closed loop Z-piezo stage movement while the sphere and cantilever-plate are in contact. However, in this geometry the direction of the force F does not coincide with the vertical movement of the Z-piezo stage (Fig 2). Thus only the normal component $F_n=F\cos(\alpha)$ is measured. In this geometry we have for the distance $d=d_0+d_{piezo}\cdot\cos(\alpha)$ and for the force $F=F_{measured}\cos(\alpha)^{-1}$. Note that in the usual case, when the sphere is attached on the cantilever, the plane can also make an angle of a few degrees from being perfectly perpendicular to the piezo stage movement. However, the correction factors become very small for a few degrees misalignment (e.g., for an angle of 2 degrees the correction factor is only 1.0006).

Finally the typical potential between a gold tip and a glass sphere was in the order of some hundreds mV, and constant within an rms noise of ~10mV over the scanned area as shown in Fig 3. Scanning surface potential imaging (SSPM) with a gold coated force modulation cantilever used in tapping mode on the spheres and gold plates indicated uniform surface potential distributions over a 1 μm$^2$ area.

**(b) Surface preparation for force measurements**

The typical BSG composition in mol is 67% O, 27% Si, 4.1% B, 0.4% Na, 1.5% Al (www.glass-warehouse.com/duran). Other manufacturers state similar values (compared to silica 67% and 33% mol for O and Si respectively). The glass substrate, on which the spheres are sintered, is beforehand cleaned by using sonic cleansing in a seven step process (each step of 5 minutes) as follows: soap->distilled water->methanol->distilled water->acetone->distilled water->nitrogen drying. The plate with the spheres are heated together for 15-30 minutes at 780 ºC after which the glass is slowly cooled down to minimize significant cracking. This procedure results in sintering of the spheres onto the clean plate.

In order to gauge whether the BSG spheres deform during heating, we have measured the mean diameter of 100 random non-sintered and sintered spheres by SEM resulting in a value of 17.7±1.7 and 18.3±1.9 μm (diameters ≈15-23μm), respectively. The latter corresponds to the certified mean sphere diameter supplied by Duke (17.3±1.4 μm with standard deviation of 2.0μm (thus from batch to batch the certified mean can vary 1.4 μm, and within a batch the standard deviation is 2.0 μm). Therefore we assume that no significant deformation took place, which would otherwise affect sphericity and local curvature. However, the sphere roughness will change (spheres become rougher for long cooling time ~12h; and smoother for shorter times ~1h). Note that sintered glass spheres on Si substrates (at 780 ºC for 2h) did not survive sonic cleaning, while the spheres on the glass sample could survive sonic cleaning at least several times. The sample with spheres is sonically cleaned with the same fluid and dried in nitrogen flow. Finally, before and after force measurements the teflon tubes and holder (with cantilever) were repeatedly flushed with the alcohol used also for force measurements. Teflon tubing was used and glue is avoided to minimize contamination. The silicone o-ring used in the AFM was sonically cleaned in ethanol. Merck ethanol (pro analysi 99.9% pure); Lab Scan methanol (analytical reagent 99.8% pure); and Lab Scan isopropanol (analytical reagent 99.7 % pure) were used. The solubility limit of KCl (Merck proanalysi salt) in ethanol and methanol is 0.034% and 0.54% mass fraction respectively [12].

**(c) Morphology analysis and separation upon contact**

The separation upon contact $d_o$ as determined from the roughness scans is $d_o$=3.5±1 nm [13]. The roughness of the BSG spheres used here is 0.36±0.15 nm rms (1.5±0.6 nm top to bottom, see Fig. 1d) and that of the cantilever 0.84±0.1 nm rms (5.4±0.7 nm top to bottom, see Fig. 1e). These results were obtained from twenty six AFM scans on different spheres and cantilevers. A large part of the scans of the spheres revealed large smooth areas (~60%). About 40% of the 800x800 nm$^2$ scans (on the spheres) had on average two features between 2 and 8 nm. Therefore, within the contact area of sphere/cantilever surface ~250x250 nm$^2$ the probability that such a feature to exist is roughly ~8 %. Areas with rare large features are ignored since for such features the measured forces will be very small. Thus force measurements on spheres, which show only very weak attraction, or no forces at all, were ignored. On the other hand isolated small nanosized features (see fig. 1d) have a tendency to deform if the surfaces are in contact so these do not affect the force measurements [13]. Indeed, for features with correlation size ~50 nm the pressures can be >100MPa when the cantilever is in contact with the sphere (pressing ~0.1-1 μm after contact to determine the deflection sensitivity depending on cantilever spring constant), which is above the ultimate strength of glass ~50 MPa. Therefore, these features do not pose a problem for the determination of the contact separation $d_o$. This is in accordance with the fact that the force measurements at short separations on different spheres were highly reproducible.

Finally, all force results were obtained for forty different spheres and ten different cantilevers. For all measurements the maximum bending of the cantilever is always a few nm upon contact, and the jump to contact was almost nonexistent, which allows us to measure surface forces down to ~4 nm separations (~$d_o$). All force curve data was corrected for the theoretical hydrodynamic force $F_h = 6\pi\eta v R^2/d$ where $\eta$, $v$ and $d$ represent respectively the liquid viscosity, approach speed, and surface separation (Fig. 4a). Also a fit is made for the zero force offset at larger separations where other forces are negligible.

## III. ELECTRICAL DOUBLE-LAYER FORCES

Figure 4a shows typical force measurements with ethanol, where every curve is an average of twenty five force curves. Before averaging all curves were shifted to the same point of contact. Moreover, since not so many points are necessary at larger separations, the noise was reduced by performing a distance-dependent averaging. The repulsive forces are double layer forces with a varying strength on different locations (spheres). Their presence can be attributed to impurity ions present in the liquid and on the sphere surface since there is a residue due to fabrication (stated also by the manufacturer). They were also obtained reproducibly with repeated measurements between different spheres showing strong and weak double layer forces. In addition, when a different cantilever is used (*in-situ*), the measured forces remained unaltered for the same sphere.

Therefore, variation of the local charging state from sphere to sphere is observable. However, on rare occasions it was found that the force was increased by consecutive repeated measuring on the same sphere (as this process was observed in particular three times). This is shown in Fig. 4b were the $1^{st}$, $5^{th}$, $10^{th}$ and $20^{th}$ successive measurements in methanol are shown. These measurements were performed within a few minutes. After observing this effect another series of twenty five measurements were performed indicating a prevailing saturated (high charge) state stable for a period of ~5 minutes, but possibly longer. Finally, in Fig. 4c we compare forces for the different alcohols. The Debye length $\lambda_D$ increases with decreasing polarity of the liquid being the highest for isopropanol and the lowest for methanol. Moreover, the solubility of salt decreases by orders of magnitude when increasing the carbon chain length, and for this reason, it is the lowest for isopropanol. By fitting an exponential function $\sim\exp(-d/\lambda_D)$ to the repulsive part, we obtain the Debye length $\lambda_D$, namely, 26±6 nm for methanol, 38±5 nm for ethanol, and 85±14 nm for isopropanol (with the standard deviation obtained from multiple measurements on different spheres).

The values of $\lambda_D$ are typical for double layer forces due to low concentration (~µmoles) of impurity ions. When solving the full non-linear Poisson-Boltzmann theory [14-16] using the solution in [16], and substituting the measured Debye lengths $\lambda_D$, the force curves can be reproduced, taking for simplicity constant potential conditions, by varying solely the surface potentials with typical values ~ -10-50 mV. For the remaining part of this article we will consider dispersion forces only for ethanol and methanol because the solubility of salt in isopropanol is rather poor, while salt ions are necessary to screen the electrostatic double layer forces.

## IV. DISPERSIVE FORCES

Figure 5a and 5b show the dielectric data for the surfaces and liquids together with calculations of the van der Waals/Casimir force using Lifshitz theory (Fig 5c) [2]. Roughness corrections to Lifshitz theory were omitted since our surfaces are smooth enough to ignore these effects at separation above 10 nm. The dielectric data for gold and ethanol/methanol were taken from [17] and [18, 19] respectively. We would like to emphasize here that a 20 nm thick gold film has optical properties which are very close to bulk systems as discussed in [17]. The force calculations from Esquivel-Sirvent [17] for the 20 nm film are similar to the results for 100-400 nm thick gold films of Svetovoy et al [17], and they fall within the scatter of that data. Indeed, skin depth effects are expected to be significant for film thickness <10 nm [17]. Therefore, we assume that the optical properties of the gold film used here are similar to those observed in [17].

The imaginary part of the measured dielectric data for the glasses (Fig 5a), as obtained from refs. [20-24], was directly integrated using the Kramers-Kronig relation to obtain the corresponding dielectric function $\varepsilon(i\zeta)$ (Fig.5b). Since for the ultraviolet (UV) spectrum (wavelength $\lambda$<300nm, $\zeta$>4eV) no data are available for soda lime or BSG, the data for pure silica were used in this range.

For gold the Drude model was used for extrapolating in the infrared/microwave (IR/MW) range ($\lambda > 30\mu m$) [17]. Alcohols are described by a three oscillator model for the dielectric function $\varepsilon(i\zeta)$ at imaginary frequencies [18, 19]

$$\varepsilon(i\zeta) = 1 + (\varepsilon_0 - \varepsilon_{IR})(1 + \zeta/\omega_{MW})^{-1} + (\varepsilon_{IR} - \eta_0^2)(1 + (\zeta/\omega_{IR})^2)^{-1} + (\eta_0^2 - 1)(1 + (\zeta/\omega_{UV})^2)^{-1}. \quad (1)$$

Here $n_o$ is the refractive index in the visible range, $\varepsilon_0$ the static dielectric constant, and $\varepsilon_{IR}$ the dielectric constant where the MW relaxation ends and the IR begins. $\omega_{MW}$, $\omega_{IR}$, and $\omega_{UV}$ are respectively the characteristic MW, IR, and UV absorption frequencies. The associated dielectric strengths and frequencies are obtained from spectral data as described in ref. [18]

It must be pointed out that there is a large spread in the dielectric data (Fig. 5a) for all glasses in the far IR range (0.001-0.1eV) [20]. However, the effect on the calculated forces is very small since the most important contribution for the force below 50 nm comes from the UV dielectric properties of glass and alcohols. This happens because the dielectric function $\varepsilon(i\zeta)$ of both glass and alcohols is relatively flat compared to $\varepsilon(i\zeta)$ for gold (Fig. 5b). In the latter case, the contribution of the IR regime on the force dominates at separations below 100 nm [17].

For all glasses the dielectric functions are very similar for frequencies above 0.1 eV (Fig 5a). Although there is no (deep) UV data available for BSG, since the UV absorption is determined by electronic processes (Si-O and other bonds, bubbles, metallic impurities, and water content) [20] and BSG consists of 80 mol% $SiO_2$, we expect its dielectric response in the UV regime to be similar with that of silica. Indeed, the transmission regime for sodalime/BSG is not very different from pure silica, i.e. by comparing the regime 270 nm - 2.7 μm to 200 nm - 3.5 μm [20]. Notably even for pure silica there is significant spread for the transmission regime. Similarly voids and impurities for evaporated gold films (whose density can vary 20% compared to bulk) can lead to variation of the dielectric functions resulting up to 15% variation in the calculated forces [17].

Scatter in the optical data in the UV regime for silica [20], leads to ~30% scatter in the calculated forces as it is shown in Fig. 6. The Lifshitz theory predicts a clear sign reversal for Au-Methanol-$SiO_2$ at 40 nm (Fig 5c) with a very weak repulsive component at separations above 40 nm. Interestingly, besides short range attraction, dispersion forces can lead to long range repulsion similar to electrostatic repulsion. However, the scaling of the repulsive part is more complex than the simpler exponential scaling of the double layer force. Because $\varepsilon(i\zeta)$ for alcohols and silica is similar, small changes in the dielectric functions of each of them will have a strong impact on the force (Figs. 5 and 6). If we use for example two different sets of optical data available for silica, the separation at which the force changes sign can increase (e.g., from 40 to 50 nm for methanol) and the force strength in the vicinity of the transition point depends crucially on this effect.

Furthermore, in Fig. 6 we show measurements for ethanol and methanol with and without added salt KCl (0.0046 and 0.013 mol/L respectively, yielding a Debye length $\lambda_D$<3 nm and therefore insignificant double layer forces). As a result only attractive van der Waals/Casimir forces were measured. The large variation found in measurements from sphere to sphere can be clarified by the uncertainty in distance and the sphere diameter. Note that most measurements of the dispersive force for different spheres are roughly within two standard deviations (~2 nm) of each other (adding further confidence for the determination of the contact separation $d_0$ from the rms roughness of cantilever and sphere surfaces). From Fig. 6a it is shown that addition of salt does on average not yield any significant difference in the measured force. The latter is roughly an order of magnitude lower than the van der Waals/Casimir force in air. Moreover, the attractive van der Waals/Casimir force was similar in magnitude for ethanol and methanol and in reasonable agreement with Lifshitz theory using the optical data in Fig 5a.

While the measured dispersive force in liquid is weak, even much weaker forces can arise for silica systems (Fig 7). Lifshitz theory predictions have been experimentally confirmed for the gold-air-gold [3-5, 13], gold-ethanol-gold [6] and in this work for gold-ethanol-glass systems. Calculations of the Casimir-Lifshitz force between two silica surfaces in ethanol based on the optical data provided here reveal forces which are even five times smaller than the forces predicted and measured here for glass-ethanol-gold [6]. The difference in force compared to the gold-air-gold system can be as large as two to three orders of magnitude. Although these weak forces should be measurable between very smooth surfaces, it will be an experimental challenge to verify the Lifshitz theory in these cases.

At last the measured force below 10 nm (Fig. 6) seems to imply the presence of other repulsive forces (resulting in weaker attraction), which appear to be relatively independent of the ionic strength and lead to deviation from the Lifshitz theory prediction. The change in scaling could be attributed to solvation effects. Below 10 nm separation also the hydrodynamic force can be influenced by surface roughness and contribute to the weaker attraction (though it is not expected to be a large effect as Fig. 4a indicates). Although, shifting the curves by only 1 nm removes the largest part of the deviations, the difference in scaling remains visible. For separations > 30 nm the discrepancy with theory and experiment is also somewhat larger since no switch in sign is observed. Besides the fact that the force switches sign at large separations where the measurement/force resolution is low (10pN), the uncertainty of the optical data (depending on fabrication conditions of the materials involved; see Fig.5a) can also explain these differences. This is because any variations of the dielectric response of glass in the UV regime would effectively shift the transition point and the force around the transition point will change by a very large factor (on a log-log scale this will look rather dramatic as can be seen in Fig 6). Furthermore the accuracy of oscillator models remains to be thoroughly investigated [25].

## V. CONCLUSIONS

Although for the gold-alcohol-glass system electrostatic double layer forces can play a dominant role at separations above 10 nm, quantum fluctuation induced dispersive forces are also present. These forces are predicted by Lifshitz theory to be weak when the dielectric function of the liquid is similar to that of one of the surfaces over all relevant frequencies. This is confirmed for first time for glass-gold surfaces immersed in alcohols, and in reasonable agreement with Lifshitz theory. Interestingly, the switch in sign of the force as predicted by Lifshitz theory for the present system was not observed. This, however, can partly be attributed to uncertainties in force measurements, because the difference between theory and experiment is on the level of the experimental noise (10 pN) in that range. On the other hand, most likely, uncertainties in theory due to scatter in the measured dielectric response of glass gold and liquids is a more serious issue The uncertainty in the theory due to scatter in the dielectric data can be even larger than the 30% uncertainty as estimated here. This originates from the use of oscillator model descriptions of the dielectric function describing the liquid [25]. For example the question arises whether the differences for the forces predicted by theory using the oscillator models is real or not. Our measurements, having 1 nm uncertainty in separation and a force resolution of 10 pN, suggest that this is not the case. Therefore, the measured forces for ethanol and methanol are almost identical over all ranges. Finally, it is shown theoretically that quantum fluctuation related dispersive forces can mimic electrostatic double layer forces having a sign switch with surface separation. While our attempts to measure this effect indicated that the sign switch was absent, it still remains an interesting phenomenon for other systems in relation, e.g., to quantum torque studies [7].

## ACKNOWLEDGMENTS

The research was carried out under project number MC3.05242 in the framework of the Strategic Research programme of the Materials innovation institute M2i (the former Netherlands Institute for Metals Research or NIMR). Financial support from the M2i is gratefully acknowledged. We would like also to acknowledge useful discussions with J. Munday, F. Capasso, V. A. Parsegian, and H. Fischer.

**FIGURE CAPTIONS**

**Figure 1 (color online) (a)** Optical image of spheres sintered on the plate **(b)** Three cantilevers together with the spheres as seen from the top in AFM. **(c)** SEM measurement of the diameter of the spheres after sintering. **(d)** Topography AFM scans of a borosilicate sphere (the stripes are noise during imaging). **(e)** Topography AFM scans of the gold coated cantilever.

**Figure 2 (color online)** Measurement setup of our inverse colloid probe system **(a)** compared to a usual colloid probe system **(b)**. Green arrows indicate the force direction in case of a sphere on a cantilever in **(b),** and in the case considered here when the cantilever acts as a plate **(a)**. Thus the plate is tilted with respect to the piezo movement in the inverse case. However, piezo movement defines the deflection sensitivity (nm/V), and thus the calibrated cantilever stiffness, which therefore is the same for both systems. Therefore, we have introduced the corrections $d=d_0+d_{piezo}\cos(\alpha)$ and $F=F_{measured}\cdot\cos(\alpha)^{-1}$.

**Figure 3 (color online)** A surface potential scan of a sphere conducted with a gold coated AFM probe in air with a lift height 40 nm. Both the surface height and potential variations are shown over a 1 $\mu m^2$ area. We performed scans on five spheres and we could not observe potential changes outside the noise level which is ~10mV. The radius of curvature of these spheres prevents us from scanning a larger area without artifacts.

**Figure 4 (color online)** Typical force measured for the indicated alcohols for different spheres at different locations. Figure (a) shows typical force measurements in ethanol on different spheres. In (b) we show a sudden change of double layer repulsion in one measurement sweep of 25 curves on a sphere. Figure (c) shows double layer forces in methanol, ethanol and isopropanol, indicating the different Debye lengths. The theoretical hydrodynamic force for the approach speed used for these measurements is also shown in Fig. 3a (dashed line) for comparison purposes.

**Figure 5 (color online) (a)** Absorption part of the dielectric function for different glasses and Gold. **(b)** Dielectric function $\varepsilon(i\zeta)$ for the materials indicated. **(c)** Lifshitz theory for glass and gold in methanol, ethanol and air. For methanol a transition from attractive to repulsive occurs at 40nm, while for ethanol the force sign changes at 100 nm.

**Figure 6 (color online) (a)** Measurements of attractive forces in ethanol, between several spheres and gold coated cantilevers: individual measurements (small red open squares), average of multiple spheres without salt when no electrostatic repulsion could be measured (red squares), and average of measurements on multiple spheres with salt (red circles). The Lifshitz theory calculation is also shown for gold-ethanol-$SiO_2$ (red solid lines, using two sets of optical data for $SiO_2$). Theory for gold and silica surfaces in air are also shown for comparison. **(b)** Average of multiple measurements on five different spheres of attractive forces measured in methanol (blue circles) with salt and compared to that of ethanol (red squares). The solid and dashed lines represent Lifshitz theory for Au-ethanol-$SiO_2$ and Au-methanol-$SiO_2$ respectively, using again two different sets of measured visible/UV optical data for $SiO_2$.

**Figure 7** Calculations for the van der Waals/Casimir force using Lifshitz theory for a sphere (18 μm diameter) and a plate for different combinations of materials. The differences in the force are very large.

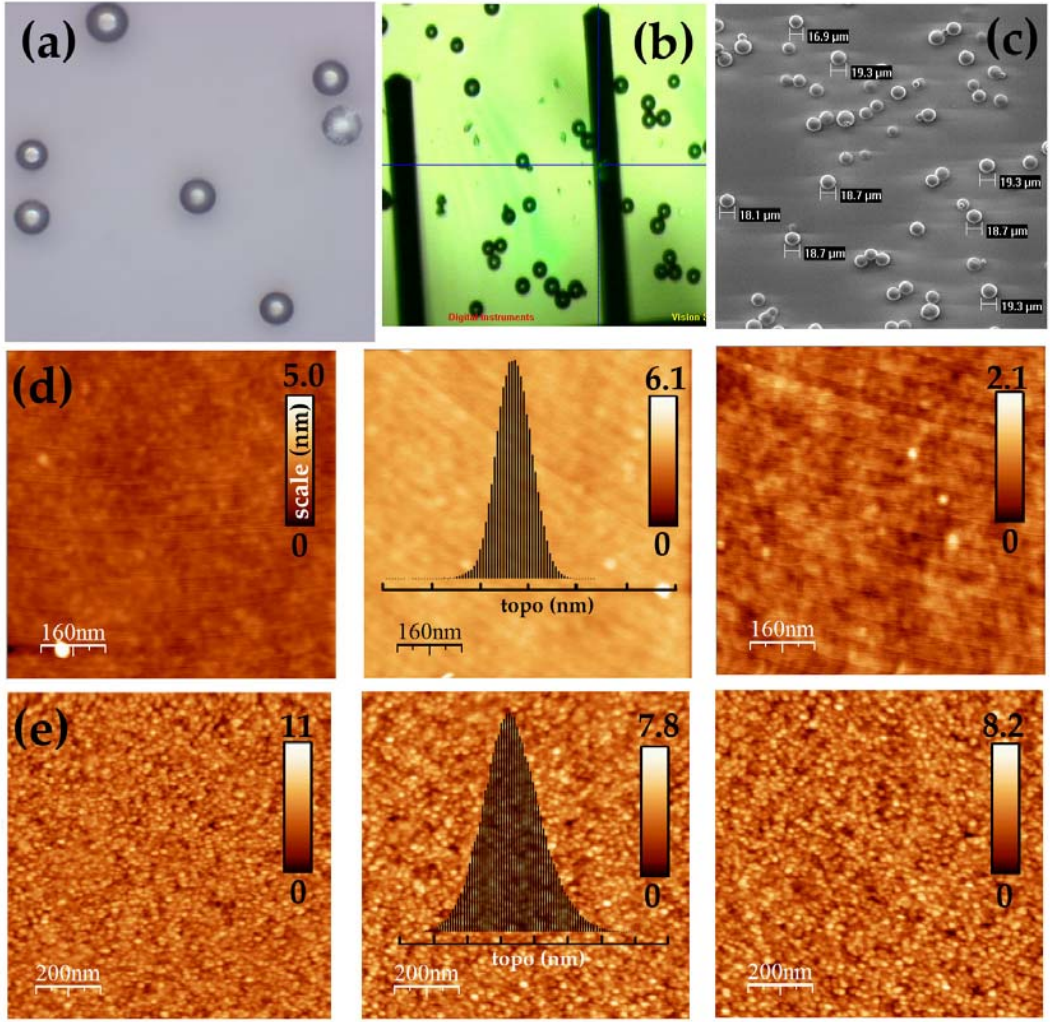

**Figure 1**

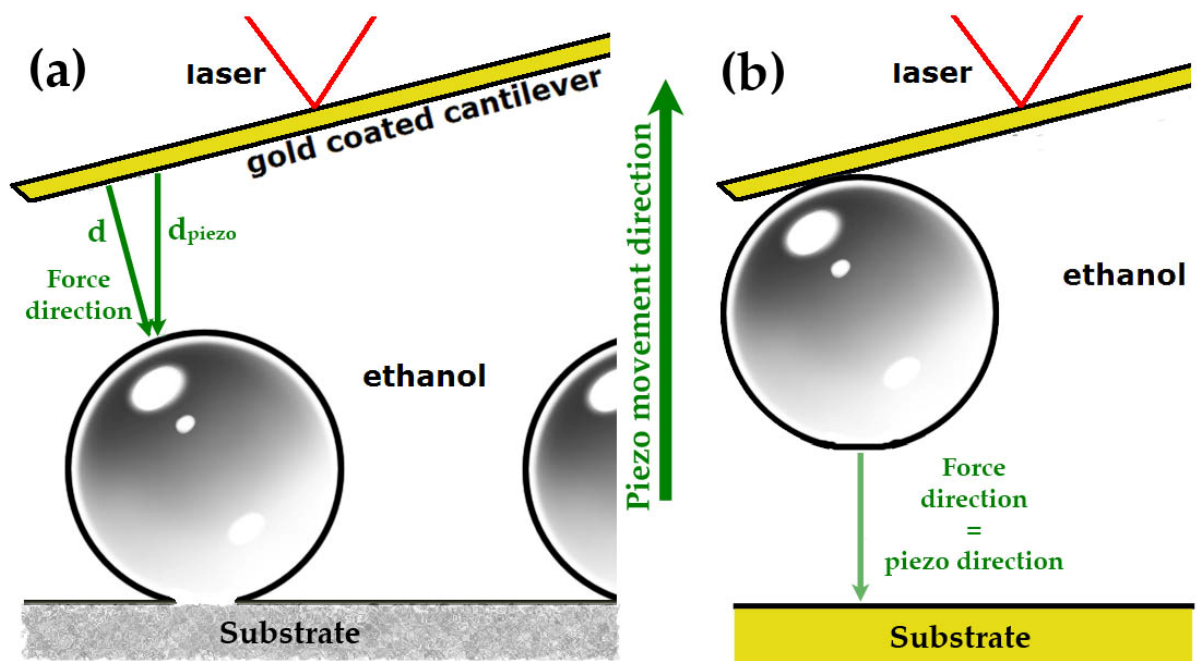

**Figure 2**

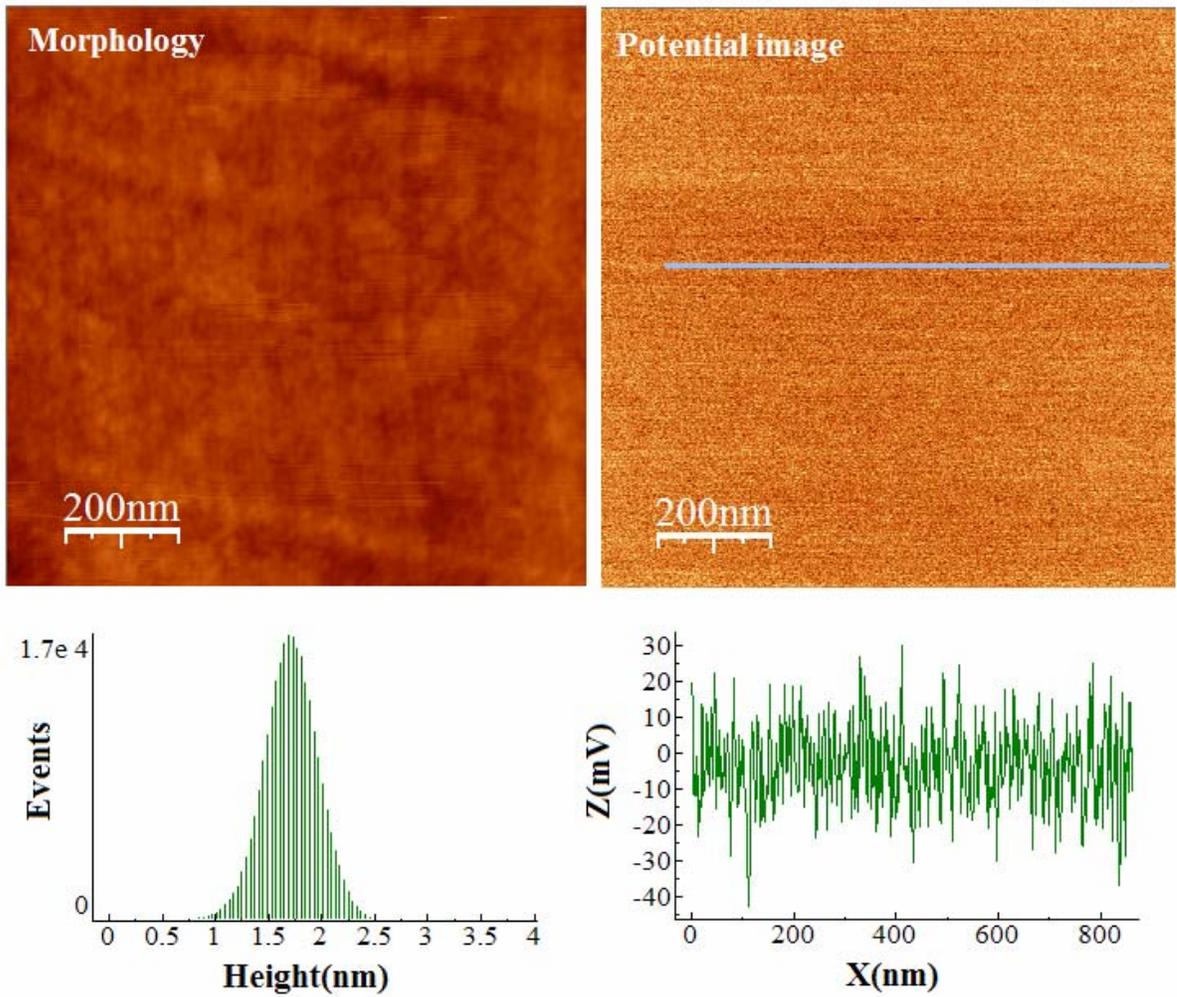

**Figure 3**

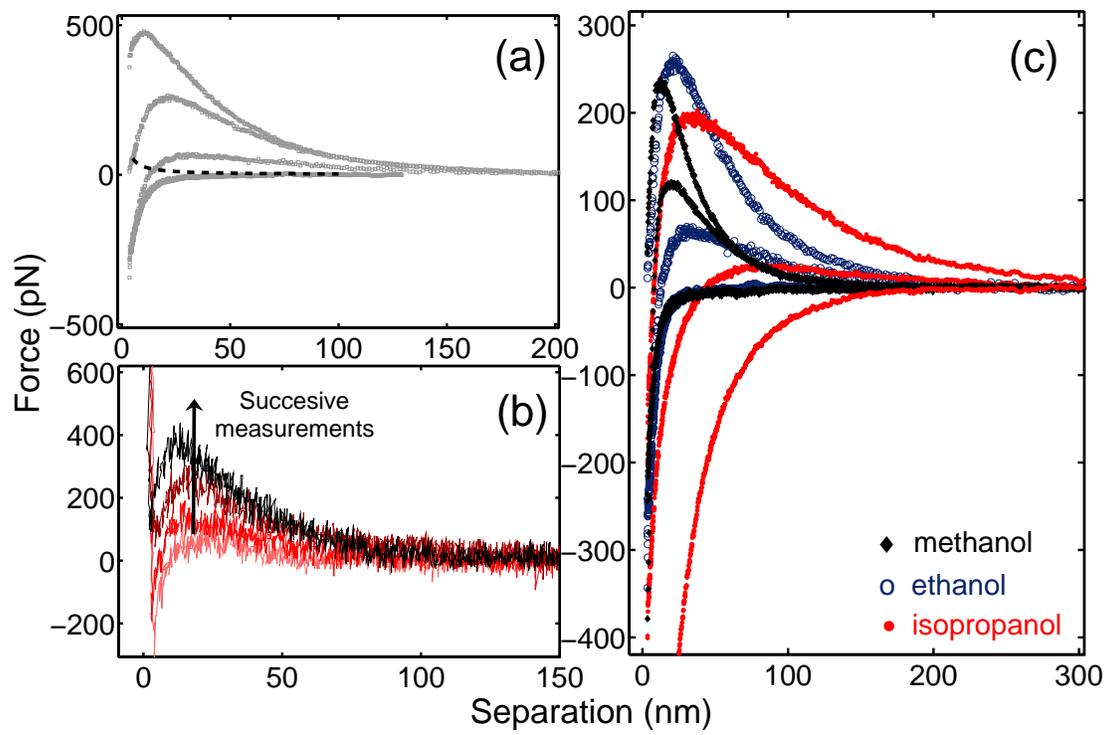

**Figure 4**

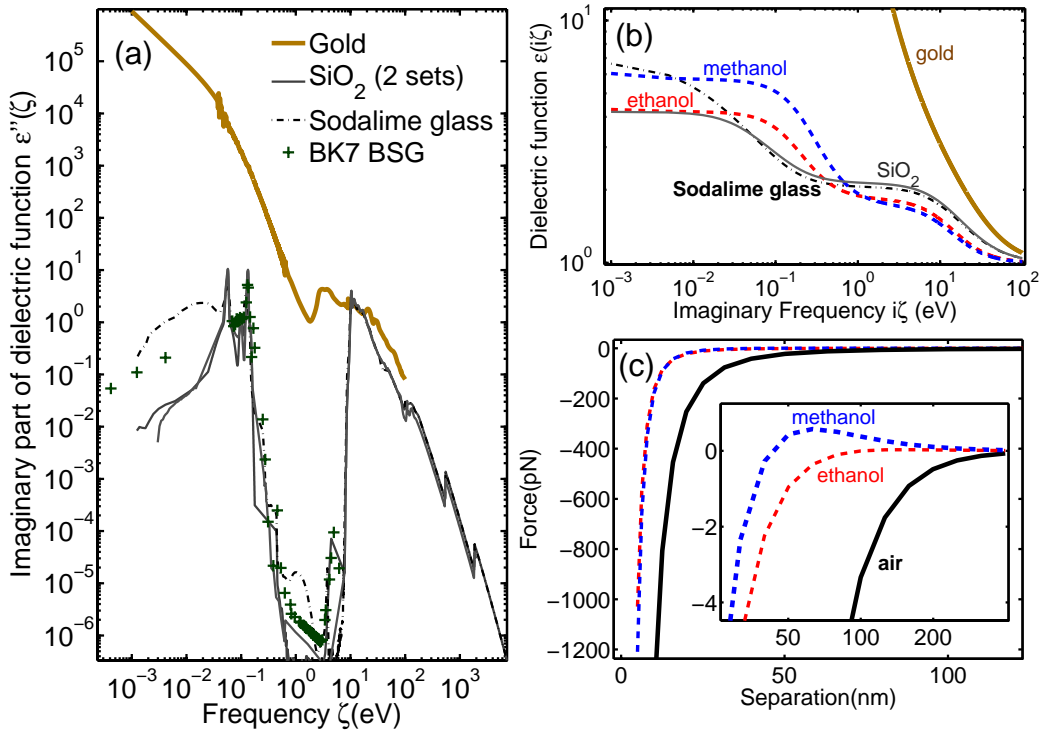

**Figure 5**

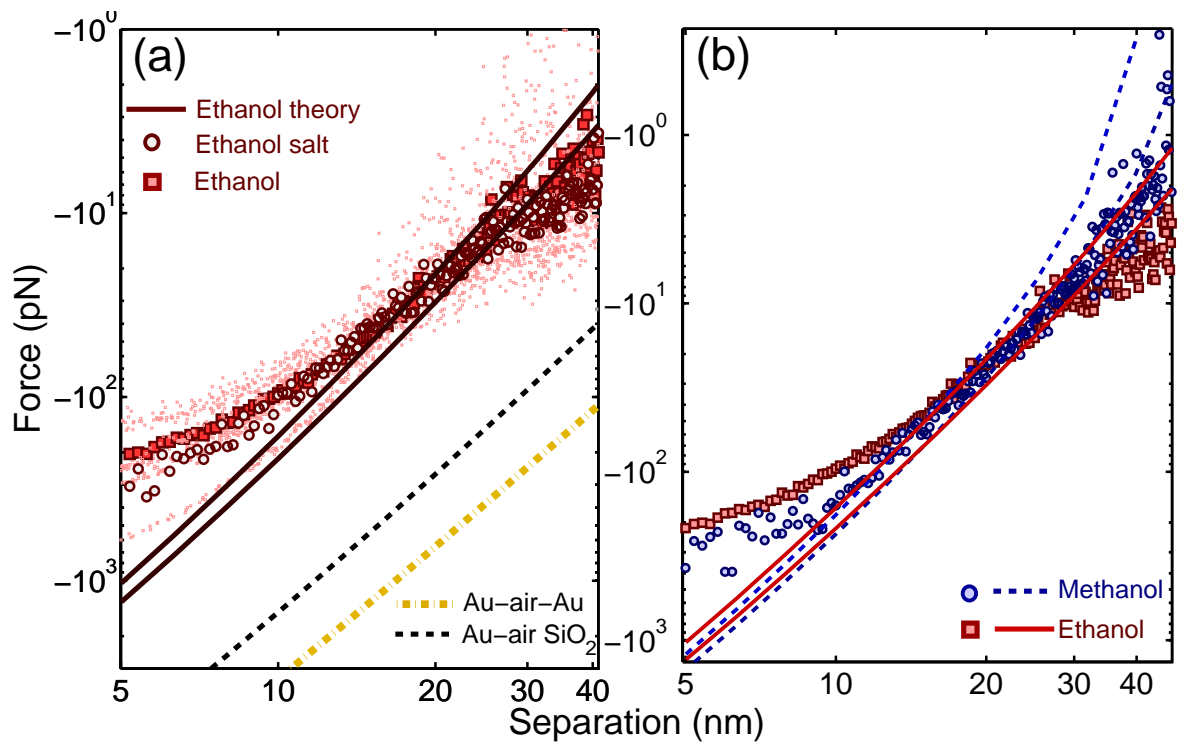

Figure 6

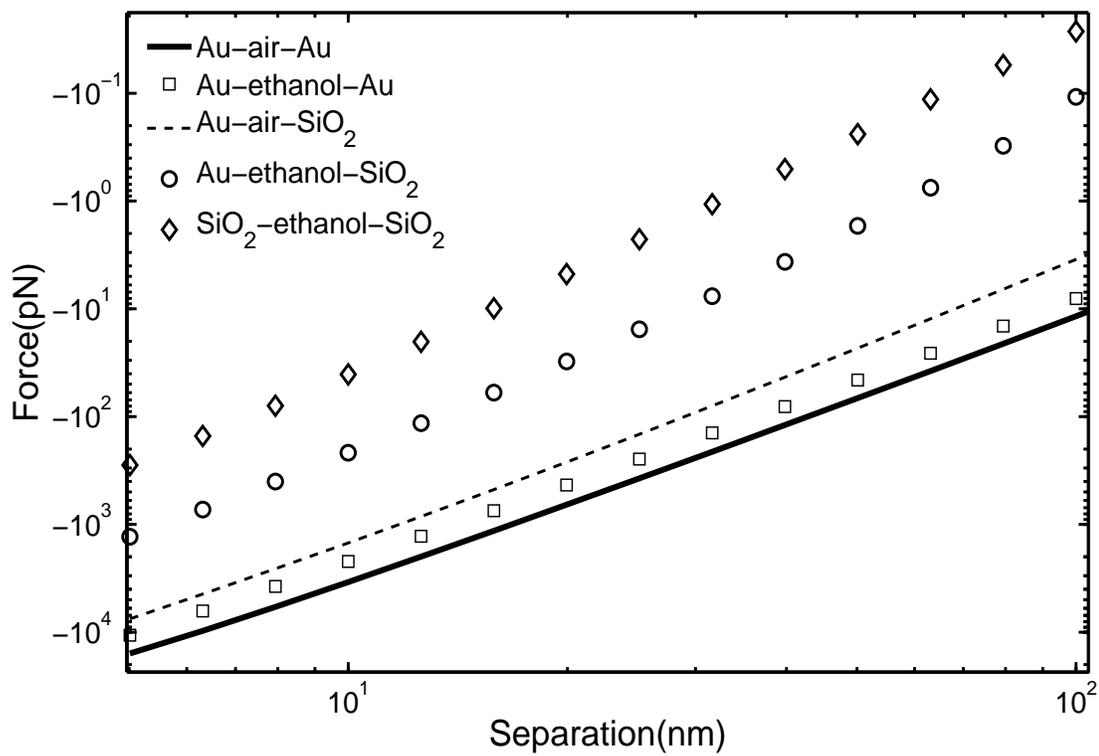

**Figure 7**